\documentclass[intlimits,twoside,a4paper]{article}

\usepackage{amsmath,amssymb}
\usepackage{graphicx,wrapfig}
\usepackage{wrapfig}

\usepackage[T2A]{fontenc}
\usepackage[cp1251]{inputenc}

\usepackage{epsf,multicol,ifthen}

\usepackage[eqsecnum]{cmpj2}

\usepackage{color}

\issue{2014}{17}{2}{23601}
\doinumber{10.5488/CMP.17.23601}

\title[Non-linear model of impurity diffusion in nanoporous materials]%
{Non-linear model of impurity diffusion in nanoporous materials upon ultrasonic treatment}
\author[R.M.~Peleshchak, O.V.~Kuzyk, O.O.~Dan'kiv]{R.M.~Peleshchak\thanks{E-mail: peleshchak@rambler.ru}
, O.V.~Kuzyk, O.O.~Dan'kiv}
\address{Drohobych Ivan Franko State Pedagogical University, 24 I.~Franko St., 82100 Drohobych, Ukraine}

\date{Received November 10, 2013, in final form February 9, 2014}
\authorcopyright{R.M.~Peleshchak, O.V.~Kuzyk, O.O.~Dan'kiv, 2014}

\begin{document}

\maketitle

\begin{abstract}
Non-linear theory of diffusion of impurities in porous materials upon ultrasonic treatment is described.
It is shown that at a defined value of deformation amplitude, an average concentration of vacancies and
temperature as a result of the effect of ultrasound possibly leads to the formation of nanoclusters of
vacancies and to their periodic educations in porous materials. It is shown that at a temperature smaller
than some critical value, a significant growth of a diffusion coefficient is observed in porous materials.
\pacs 66.30.Lw, 43.35.Fj
\end{abstract}

\vspace{5mm}

\section{Introduction}
It is known that in an ultrasonically treated solid, the concentration of defects, in particular
the concentration of vacancies nonlinearly depends on both temperature and acoustic vibration
intensity~\cite{Bul08,Abr94}. Moreover, in a certain ultrasonic and temperature range, one can
observe a significant increase (more than by one order) of the defect structure of a sample, i.e.,
the acoustic effect is clearly synergetic in this case. In work~\cite{Bul08} it has been demonstrated
that the equilibrium vacancy concentration can be high even at low temperatures, if the bulk deformation
exceeds some critical value. The processes of self-organization of vacancies (that interact with one
another and with the crystalline matrix through the deformation field) into separate clusters and
periodic structures is possible if their concentration is high enough~\cite{Eme96}. The formation of
a periodic pore lattice in metal and in dielectric materials irradiated with high-energy neutron and
electron beams has been observed in works~\cite{Sik74,Cha76}.

In works~\cite{Zav02,Oli02,Ost95} it has been demonstrated that by means of a supersonic wave one can
control the transport properties of semiconductors and change their structure due to the processes of
impurity atom diffusion, dissolution and the formation of complexes, as well as the formation of
impurity atom clusters and intrinsic defects in periodic deformation fields.

Extrinsic heterogeneous deformation causes the change of point defect chemical potential and leads to
the directional diffusion flows. In work~\cite{Ost02} it has been empirically determined that Si
ultrasonic processing can stimulate a diffusion at room temperature. A significant increase (more
than $2\div3$ times larger) of carbon diffusion coefficient in steel is observed at a fixed
temperature in a certain range of acoustic vibration amplitude~\cite{Kul78}.

In this work, a nonlinear diffusion deformation theory of vacancy cluster formation in ultrasonically
treated porous material is made, in order to specify the ultrasonic effect on the impurity diffusion coefficient.

\clearpage

\section{The model}

\begin{wrapfigure}{i}{0.5\textwidth}
%\begin{figure}
\centerline{
\includegraphics[width=0.4\textwidth]{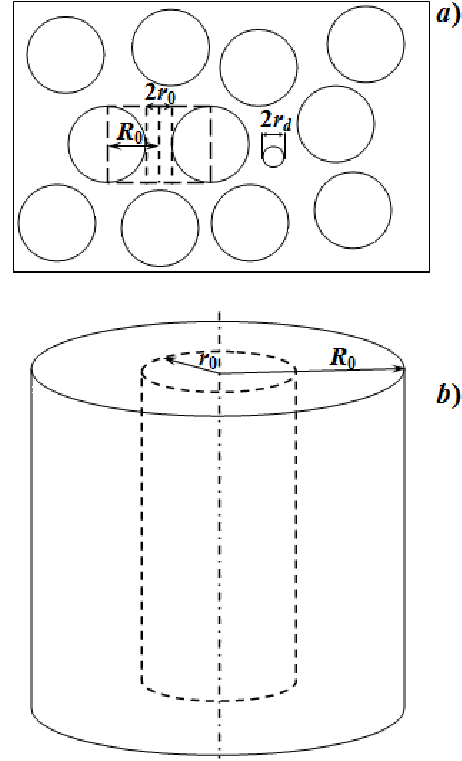}
}
\caption{Geometrical model of a porous medium with impurity.}
\label{fig1}
%\end{figure}
\vspace{-25mm}
\end{wrapfigure}

We model a porous medium using a system of spherical particles (granules),
the radius of which is
$r_\mathrm{g} = R_0 - r_0$, and the radius of impurities is $r_\mathrm{d}$ (figure~\ref{fig1}~(a)),
where $2r_0$ is the space between the granules, considered
herein below as the pore diameter
(figure~\ref{fig1}~(b)). We select a cylindrical bulk element of a porous material (figure~\ref{fig1})
with the radius $R_0$ and the cylindrical pore having the radius $r_0$ ($R_0 \gg r_0$).
The average concentration of vacancies of this system is $N_{\mathrm{d}0}$.

Considering the nonlocal Hooke's law \cite{Kun70}, the energy of vacancy interaction with
matrix atoms $U_\mathrm{d}^{{\mathrm{int}} }$  through the elastic field can be determined as follows:
\begin{equation}
\label{eq1}
U_\mathrm{da}^{{\mathrm{int}} } (r)
=  - \int {\lambda \left( {\left| {r' - r} \right|} \right)\varepsilon (r')\Delta \Omega_\mathrm{d} } \rd r',
\end{equation}
where  $\lambda$ is the elastic moduli operator \cite{Eme96}. Introducing the variable
$\tau  = r' - r$ and expanding $\varepsilon (r + \tau )$ in a Taylor series by $\tau$, we obtain:
\begin{align}
\label{eq2}
U_\mathrm{da}^{{\mathrm{int}} } (r) &=  - \int {\lambda \left( {\left| \tau  \right|} \right)\varepsilon
(r + \tau )\Delta \Omega_\mathrm{d} } \rd\tau\nonumber\\
&=  - \int {\lambda \left( {\left| \tau  \right|} \right)\left[ {\varepsilon (r) + \frac{{\partial ^2
\varepsilon (r)}}{{\partial r^2 }}\frac{{\tau ^2 }}{2}} \right]\Delta \Omega_\mathrm{d} } \rd\tau  \nonumber\\
&=  - K\varepsilon (r)\Delta \Omega_\mathrm{d}  - K\frac{{\partial ^2 \varepsilon (r)}}{{\partial r^2 }}
r_\mathrm{da}^2 \Delta \Omega_\mathrm{d} \,,
\end{align}
where $K = \int {\lambda \left( {\left| \tau  \right|} \right)\rd\tau }$ is an elasticity coefficient \cite{Eme96};
$r_{\mathrm{da}} =  \left[\frac{1}{2}{\int {\lambda \left( {\left| \tau  \right|} \right)\tau ^2 \rd\tau } }\big/
{\int {\lambda \left( {\left| \tau  \right|} \right)\rd\tau } }\right]^{1/2}$
is the characteristic length of interaction between the vacancy and matrix atoms; $\Delta \Omega_\mathrm{d}$
is the crystalline volume change caused by one defect; $\varepsilon (r)$ is a strain tensor radial component.
The elastic field of a solid acts on the vacancy with the force:
\begin{align}
\label{eq3}
\vec F &=  - \textrm{grad}\,U_\mathrm{da}^{{\mathrm{int}} }  \nonumber\\
&= K\textrm{grad}\left[ {\varepsilon (r)\Delta \Omega_\mathrm{d}  + \frac{{\partial ^2 \varepsilon (r)}}{{\partial r^2 }}
r_\mathrm{da}^2 \Delta \Omega_\mathrm{d} } \right].
\end{align}
Under this force, the defects in the elastic field get the velocity
\begin{equation}
\label{eq4}
\vec \upsilon  = \mu \vec F = \frac{{D_\mathrm{v}  K}}{{k_\mathrm{B} T}}\textrm{grad}
\left[ {\varepsilon (r)\Delta \Omega_\mathrm{d}  + \frac{{\partial ^2 \varepsilon (r)}}{{\partial r^2 }}
r_\mathrm{da}^2 \Delta \Omega_\mathrm{d} } \right],
\end{equation}
where $\mu$, $D_\mathrm{v}$  are, respectively, the mobility and vacancy diffusion coefficient; $T$ is the
temperature; $k_\mathrm{B}$ is the Boltzmann constant. Here, we use the Einstein relation to determine the impurity mobility.

As we can see from (\ref{eq4}), the vacancy velocity in the elastic field is determined by the deformation
gradients and crystal volume gain due to these defects. Thus, the defects that are the compression centers
($\Delta\Omega_d < 0$), in particular the vacancies, will move to the area of the relative compression
[the direction of the velocity vector of vacancy coincides with the direction of the vector $ -\textrm{grad}\,\varepsilon (r)$].

Taking (\ref{eq4}) into account, the stationary flow of vacancies can be presented as
follows:
\begin{equation}
\label{eq5}
j =  - D_\mathrm{v}  \frac{{\partial N_\mathrm{d} }}{{\partial r}} + N_\mathrm{d} \frac{{D_\mathrm{v}  K}}
{{k_\mathrm{B} T}}\left( {\frac{{\partial \varepsilon }}{{\partial r}}\Delta \Omega_\mathrm{d}
+ \frac{{\partial ^3 \varepsilon }}{{\partial r^3 }}r_\mathrm{da}^2 \Delta \Omega_\mathrm{d} } \right).
\end{equation}

The potential energy density of the elastic defect-free continuum that takes into account the anharmonic
components can be presented as follows:
\begin{equation}
\label{eq6}
U_\mathrm{a}  = \frac{1}{2}E\varepsilon ^2 \left( r \right) + \frac{1}{3}E\alpha \varepsilon ^3 \left( r \right)
+ \frac{1}{4}E\beta \varepsilon ^4 \left( r \right) + Ea_0^2 \frac{{\partial ^2 \varepsilon
\left( r \right)}}{{\partial r^2 }}\varepsilon \left( r \right),
\end{equation}
where $E$ is the modulus of elasticity;  $\alpha$, $\beta$  are the elastic anharmonicity
constants; $a_0$ is the characteristic distance of the crystalline matrix atom interaction
that is roughly equal to the matrix lattice parameter.

Then, taking into account (\ref{eq2}) and (\ref{eq6}), the free energy density of a crystal
with vacancies can be presented as follows:
\begin{align}
\Phi  = U_\mathrm{a}  + N_\mathrm{d} U_\mathrm{d}^{{\mathrm{int}} }  - TS = & \frac{1}{2}E\varepsilon ^2
\left( r \right) + \frac{1}{3}E\alpha \varepsilon ^3 \left( r \right) + \frac{1}{4}E\beta \varepsilon ^4
\left( r \right) + Ea_0^2 \frac{{\partial ^2 \varepsilon \left( r \right)}}{{\partial r^2 }}\varepsilon
\left( r \right) \nonumber
\\\label{eq7}
 &- KN_\mathrm{d} (r)\varepsilon (r)\Delta \Omega_\mathrm{d}^{}  - KN_\mathrm{d} (r)\frac{{\partial ^2
 \varepsilon (r)}}{{\partial r^2 }}r_\mathrm{da}^2 \Delta \Omega_\mathrm{d}^{}  - TS,
\end{align}
where $S$ is the entropy density.

Applying the relation  $\sigma  = {{\partial \Phi }}/{{\partial \varepsilon }}$, we obtain the mechanical stress expression:
\begin{equation}
\label{eq8}
\sigma  = E\varepsilon  + E\alpha \varepsilon ^2  + E\beta \varepsilon ^3
+ Ea_0^2 \frac{{\partial ^2 \varepsilon }}{{\partial r^2 }} - KN_\mathrm{d} \Delta \Omega_\mathrm{d} \, .
\end{equation}

The mechanical stress in an ultrasonically treated solid, subjected to the anharmonic components, is as follows:

\begin{equation}
\label{eq9}
\tilde \sigma  = \sigma  + E\varepsilon _0 \cos \omega t + E\alpha \left( {\varepsilon _0 \cos \omega t} \right)^2
+ E\beta \left( {\varepsilon _0 \cos \omega t} \right)^3 ,
\end{equation}
where $\varepsilon_0$ is an ultrasonically induced deformation amplitude. Here, the wavelength  $\lambda \gg R_0$.
Having averaged by the time, we obtain:
\begin{align}
\tilde \sigma  &= E\varepsilon  + E\alpha \varepsilon ^2 \left(1 + \frac{{\varepsilon _0^2 }}{{2\varepsilon ^2 }}\right)
+ E\beta \varepsilon ^3  + Ea_0^2 \frac{{\partial ^2 \varepsilon }}{{\partial r^2 }} - KN_\mathrm{d} \Delta \Omega_\mathrm{d}
\nonumber\\
\label{eq10}
 &= E\varepsilon  + E\tilde \alpha \varepsilon ^2  + E\beta \varepsilon ^3  + Ea_0^2
 \frac{{\partial ^2 \varepsilon }}{{\partial r^2 }} - KN\Delta \Omega_\mathrm{d}\,,
\end{align}
where  $\tilde \alpha  = \alpha \left(1 + {{\varepsilon _0^2 }}/{{2\varepsilon ^2 }}\right)$.

Let us present the vacancy concentration and deformation as follows:
\begin{equation}
\label{eq11}
N_\mathrm{d} (r) = N_1 (r) + N_0 \,,
\end{equation}
\begin{equation}
\label{eq12}
\varepsilon (r) = \varepsilon _1 (r) + N_0 \Delta \Omega_\mathrm{d} +  \varepsilon _0 {\rm }\textrm{cos}\omega t \,
,
\end{equation}
where  $N_1 (r)$, $\varepsilon _1 (r)$   are the space inhomogeneous components
of vacancy concentration and deformation, respectively;   $N_1 (r) \ll N_0$. Thus,
\[
\tilde \alpha  = \alpha \left[ {1 + \frac{{\varepsilon _0^2 }}{{2\left( {N_0 \Delta \Omega_\mathrm{d} } \right)^2  + \varepsilon _0^2 }}} \right].
\]

From the strained solid equilibrium condition ${{\partial \tilde \sigma }}/{{\partial r}} = 0$,
we obtain the following deformation equation:
\begin{equation}
\label{eq13}
E\frac{{\partial \varepsilon }}{{\partial r}} + E\tilde \alpha \frac{{\partial \left( {\varepsilon ^2 } \right)}}{{\partial r}}
+ E\beta \frac{{\partial \left( {\varepsilon ^3 } \right)}}{{\partial r}} + Ea_0^2 \frac{{\partial ^3
\varepsilon }}{{\partial r^3 }} - K\frac{{\partial N_\mathrm{d} }}{{\partial r}}\Delta \Omega_\mathrm{d}  = 0.
\end{equation}

Taking into account (\ref{eq5}), the diffusion steady-state equation for vacancies can be written as follows:
\begin{equation}
\label{eq14}
{\rm div}\left[ {D_\mathrm{v}  \frac{{\partial N_\mathrm{d} }}{{\partial r}}
- N_\mathrm{d} \frac{{D_\mathrm{v}  K}}{{k_\mathrm{B} T}}\left( {\frac{{\partial \varepsilon }}{{\partial r}}\Delta \Omega_\mathrm{d}
+ \frac{{\partial ^3 \varepsilon }}{{\partial r^3 }}r_\mathrm{da}^2 \Delta \Omega_\mathrm{d} } \right)} \right]
+ G_\mathrm{d}  - \frac{{N_\mathrm{d} }}{{\tau_\mathrm{d} }} = 0,
\end{equation}
where  $G_\mathrm{d}$, $\tau_\mathrm{d}$  are the generation rate and vacancy lifetime, respectively.

To find the vacancy concentration distribution and deformation in the investigated structure,
one needs to solve the system of nonlinear differential equations (\ref{eq13}) and (\ref{eq14}).
Substituting (\ref{eq11}), (\ref{eq12}) into (\ref{eq13}), (\ref{eq14}), and taking into account that
$N_1 (r) \ll N_0$ and at $r \to R_0$,  the conditions ${{\partial N_1 }}/{{\partial r}} = 0$  and
${{\partial \varepsilon _1 }}/{{\partial r}} = 0$ must be kept, we obtain that $N_0  = G_\mathrm{d} \tau_\mathrm{d}$,
and the equation for $N_1(r)$ and  $\varepsilon_1(r)$ reads as follows:
\begin{equation}
\label{eq15}
D_\mathrm{v}  \frac{{\partial N_1 }}{{\partial r}} - N_0 \frac{{D_\mathrm{v}  \theta_\mathrm{v} }}
{{k_\mathrm{B} T}}\left( {\frac{{\partial \varepsilon _1 }}{{\partial r}} + \frac{{\partial ^3
\varepsilon _1 }}{{\partial r^3 }}r_\mathrm{da}^2 } \right) = 0,
\end{equation}
\begin{equation}
\label{eq16}
\frac{{\partial \varepsilon _1 }}{{\partial r}} + \tilde \alpha \frac{{\partial \left( {\varepsilon _1^2 } \right)}}
{{\partial r}} + \beta \frac{{\partial \left( {\varepsilon _1^3 } \right)}}{{\partial r}} + a_0^2 \frac{{\partial ^3
\varepsilon _1 }}{{\partial r^3 }} - \frac{{\theta _\mathrm{v}  }}{E}\frac{{\partial N_1 }}{{\partial r}} = 0,
\end{equation}
where $\theta_\mathrm{v}  = K\Delta \Omega_\mathrm{d}$  is the vacancy deformation potential.

Integrating the equation (\ref{eq15}), we obtain:
\begin{equation}
\label{eq17}
N_1  = \frac{{N_0 \theta_\mathrm{v} }}{{k_\mathrm{B}T}}\left( {\varepsilon _1
+ \frac{{\partial ^2 \varepsilon _1 }}{{\partial r^2 }}r_\mathrm{da}^2 } \right).
\end{equation}

Substituting (\ref{eq17}) into (\ref{eq16}), we obtain the deformation equation that, after integration, can be written as follows:
\begin{equation}
\label{eq18}
\frac{{\partial ^2 \varepsilon _1 }}{{\partial r^2 }} - a\varepsilon _1  + f\varepsilon _1^2  - c\varepsilon _1^3  = 0,
\end{equation}
where
\[
a = \frac{{1 - \frac{{N_0 }}{{N_\mathrm{c} }}}}{{r_\mathrm{da}^2 \left( {\frac{{N_0 }}
{{N_\mathrm{c} }} - \frac{{a_0^2 }}{{r_\mathrm{da}^2 }}} \right)}}\,,
\qquad
f = \frac{{\left| {\tilde \alpha } \right|}}{{r_\mathrm{da}^2 \left( {\frac{{N_0 }}{{N_\mathrm{c} }}
- \frac{{a_0^2 }}{{r_\mathrm{da}^2 }}} \right)}}\,,
\qquad
c = \frac{\beta }{{r_\mathrm{da}^2 \left( {\frac{{N_0 }}{{N_\mathrm{c} }} - \frac{{a_0^2 }}{{r_\mathrm{da}^2 }}} \right)}}\,,
\qquad
N_\mathrm{c}  = \frac{{E  k_\mathrm{B} T}}{{\theta_\mathrm{v}^2 }}\,.
\]
Here, we have taken into account that   $\alpha < 0$,   $\beta > 0$  \cite{Eme96}.

\section{Formation of vacancy nanoclusters and their periodic structures}

The solution of the equation (\ref{eq18}) is as follows:
\[
r = \int {\frac{{\rd\varepsilon _1 }}{{\sqrt {a\varepsilon _1^2
- \frac{{2f\varepsilon _1^3 }}{3} + \frac{{c\varepsilon _1^4 }}{2}} }}}  + r_\mathrm{c} \,,
\]
where $r_\mathrm{c}$ is the constant of integration.

Making a substitution  $\varepsilon _1  = \frac{1}{z}$, one can present this integral as
follows:
\begin{equation}
\label{eq19}
r - r_\mathrm{c}  =  - \int {\frac{{\rd z}}{{\sqrt {a\left( {z - \frac{f}{{3a}}} \right)^2  + \Delta } }}} \,,
\end{equation}
where   $\Delta  =  - {{f^2 }}/{{(9a)}} + {c}/{2}$.

The integral (\ref{eq19}) is expressed by analytic functions whose type we determine by the sign of the coefficients $a$ and  $\Delta$.

If the following conditions are fulfilled:
\begin{equation}
\label{eq20}
\frac{{N_0 }}{{N_\mathrm{c} }} < \frac{{a_0^2 }}{{r_\mathrm{da}^2 }}  \qquad  \text{and}
\qquad \frac{{2\tilde \alpha ^2 }}{{9\beta }} < 1 - \frac{{N_0 }}{{N_\mathrm{c} }} \qquad
(a<  0 \quad \text{and} \quad \Delta <  0),
\end{equation}
then  $\varepsilon_1 = 0$, and  $N(r) = N_0 {\rm   }$, the distribution of vacancies is spatially homogeneous.
Taking into account that ${{2\alpha ^2 }}/({{9\beta }}) = {4}/{9}$ \cite{Eme96} and
$\tilde \alpha  = \alpha \left\{ {1 + {{\varepsilon _0^2 }}\big/\left[{{2\left( {N_0 \Delta \Omega_\mathrm{d} } \right)^2
+ \varepsilon _0^2 }}\right]} \right\}$, the conditions (\ref{eq20}) can be written as follows:
\begin{equation}
\label{eq21}
\frac{{N_0 }}{{N_\mathrm{c} }} < \frac{{a_0^2 }}{{r_\mathrm{da}^2 }}\qquad \text{and} \qquad
\varepsilon _0^2  < \left( {\frac{{\sqrt 2 k_\mathrm{B} T}}{{\theta_\mathrm{v} }}} \right)^2
\left( {\frac{{N_0 }}{{N_\mathrm{c} }}} \right)^2
\left( {\frac{1}{{\frac{3}{2}\sqrt {1 - \frac{{N_0 }}{{N_\mathrm{c} }}}  - 1}} - 1} \right)^{ - 1} .
\end{equation}

If the average defect concentration exceeds the value  $N_0  = N_\mathrm{c} {{a_0^2 }}\big/{{r_\mathrm{da}^2 }}$,
whatever is the supersonic wave deformation amplitude, the spatially nonuniform solution becomes unstable,
and there appears a new spatially nonuniform stationary state (i.e., the formation of clusters or periodic
vacancy structures). Moreover, if the second condition in (\ref{eq21}) is not fulfilled, vacancy clusters
will always appear. If the concentration of clusters or periodic vacancy structures is constant, then their
formation conditions depend on the temperature. In particular, the conditions (\ref{eq21}) can be written as follows:
\begin{equation}
\label{eq22}
\frac{{T_\mathrm{c} }}{T} < \frac{{a_0^2 }}{{r_\mathrm{da}^2 }}\, , \qquad
\varepsilon _0^2  < \left( {\frac{{\sqrt 2 k_\mathrm{B} T_\mathrm{c} }}{{\theta_\mathrm{v} }}} \right)^2
\left( {\frac{1}{{\frac{3}{2}\sqrt {1 - \frac{{T_\mathrm{c} }}{T}}  - 1}} - 1} \right)^{ - 1} ,
\end{equation}
where   $T_\mathrm{c}  = {{\theta_\mathrm{v}^2 N_0 }}/({{E  k_\mathrm{B} }})$.

In other cases, depending on the values $N_0 (T)$ and  $\varepsilon_0$, the solution of the equation (\ref{eq18}) will be as follows:
\begin{itemize}
  \item $a > 0$   and   $\Delta > 0$:
\[\varepsilon _1 \left( r \right) =  - \frac{A}{{B + \textrm{sh}\left[ { - \sqrt a \left( {r - r_0 } \right)} \right]}} \qquad
 \text{at} \qquad  \frac{{a_0^2 }}{{r_\mathrm{da}^2 }} < \frac{{N_0 }}{{N_\mathrm{c} }} < 1 - \frac{{2\tilde \alpha ^2 }}{{9\beta }}
 \qquad \text{and}\]
 \begin{equation}
 \label{eq23}
 \varepsilon _0^2  < \left( {\frac{{\sqrt 2 k_\mathrm{B} T}}{{\theta_\mathrm{v} }}} \right)^2
 \left( {\frac{{N_0 }}{{N_\mathrm{c} }}} \right)^2 \left( {\frac{1}{{\frac{3}{2}
 \sqrt {1 - \frac{{N_0 }}{{N_\mathrm{c} }}}  - 1}} - 1} \right)^{ - 1} ;
\end{equation}
\item $a > 0$   and   $\Delta < 0$:
\[
\varepsilon _1 \left( r \right) =  - \frac{A}{{B + \textrm{ch}\left[ {\sqrt a \left( {r - r_0 } \right)} \right]}}
\qquad  \text{at} \qquad 1 - \frac{{2\tilde \alpha ^2 }}{{9\beta }} < \frac{{N_0 }}{{N_\mathrm{c} }} < 1 \qquad \text{and}
\]
\[
\varepsilon _0^2  < \left( {\frac{{\sqrt 2 k_\mathrm{B} T}}{{\theta_\mathrm{v} }}} \right)^2
\left( {\frac{{N_0 }}{{N_\mathrm{c} }}} \right)^2 \left( {\frac{1}{{\frac{3}{2}
\sqrt {1 - \frac{{N_0 }}{{N_\mathrm{c} }}}  - 1}} - 1} \right)^{ - 1} ,\]
or at
\begin{equation}
\label{eq24}
\frac{{a_0^2 }}{{r_\mathrm{da}^2 }} < \frac{{N_0 }}{{N_\mathrm{c} }} < 1
- \frac{{2\tilde \alpha ^2 }}{{9\beta }}, \qquad \varepsilon _0^2  >
\left( {\frac{{\sqrt 2 k_\mathrm{B} T}}{{\theta_\mathrm{v} }}} \right)^2
\left( {\frac{{N_0 }}{{N_\mathrm{c} }}} \right)^2 \left( {\frac{1}{{\frac{3}{2}
\sqrt {1 - \frac{{N_0 }}{{N_\mathrm{c} }}}  - 1}} - 1} \right)^{ - 1} ;
\end{equation}
\item $a < 0$   and   $\Delta > 0$:

\[
\varepsilon _1 \left( r \right) =  - \frac{A}{{B + \sin \left[ {\sqrt {\left| a \right|} \left( {r - r_0 } \right)} \right]}}
\qquad \text{at} \qquad \frac{{N_0 }}{{N_\mathrm{c} }} > 1,\]
or at
\begin{equation}
\label{eq25}
\varepsilon _0^2  > \left( {\frac{{\sqrt 2 k_\mathrm{B} T}}{{\theta_\mathrm{v} }}} \right)^2
\left( {\frac{{N_0 }}{{N_\mathrm{c} }}} \right)^2 \left( {\frac{1}{{\frac{3}{2}
\sqrt {1 - \frac{{N_0 }}{{N_\mathrm{c} }}}  - 1}} - 1} \right)^{ - 1}
\qquad \text{and} \qquad \frac{{N_0 }}{{N_\mathrm{c} }} < \frac{{a_0^2 }}{{r_\mathrm{da}^2 }}\,,
\end{equation}
where  $A = 3\sqrt 2 \left| a \right|\left( {\left| {9ca  -  2f^2 } \right|} \right)^{ - {\rm  }\frac{1}{2}}$,
$B = \sqrt 2 f\left( {\left| {9ca  -  2f^2 } \right|} \right)^{ - {\rm  }\frac{1}{2}}$.
\end{itemize}

The constant of integration $r_\mathrm{c}$ is chosen for the reason that the maximum vacancy clustering occurs at the pore surface,
so that the condition  $r_\mathrm{c} = r_0$ is fulfilled.

\begin{figure}[!t]
\centerline{
\includegraphics[width=0.55\textwidth]{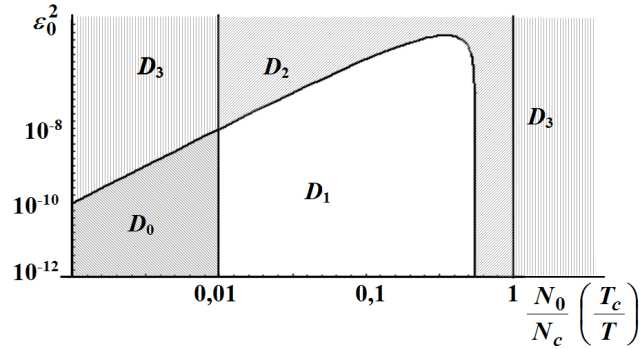}
}
\caption{Areas of possible vacancy cluster formation depending on the values
                    $\varepsilon _0^2$  and ${{N_0 }}/{{N_\mathrm{c} }}\left( {{{T_\mathrm{c} }}/{T}} \right)$:
$D_0$~--- the distribution of vacancies is spatially homogeneous;
$D_1$~--- there appears an asymmetric vacancy cluster [formula (\ref{eq23})];
$D_2$~--- there appears a symmetric vacancy cluster [formula (\ref{eq24})];
$D_3$~--- there appears a periodic vacancy lattice [formula (\ref{eq25})].\label{fig2}}
\end{figure}

Thus, vacancy clusters or their periodic structures are formed at certain values of the defect concentration $N_0$
(or the temperature $T$) and supersonic wave amplitude  $\varepsilon_0$. In figure~\ref{fig2}, the areas of possible
formation of vacancy clusters depending on the values  $\varepsilon _0^2$ and
${{N_0 }}/{{N_\mathrm{c} }}\left( {{{T_\mathrm{c} }}/{T}} \right)$   are plotted.
The calculations are made for the following parameter values:  ${{a_0^2 }}\big/{{r_\mathrm{da}^2 }} = 0.01$;
${{\sqrt 2 k_\mathrm{B} T}}\big/{{\theta_\mathrm{v} }} = 0.01$. At the average defect concentration $N_0 > N_\mathrm{c}$
(or the temperature $T < T_\mathrm{c}$) in porous material, periodic defect-deformation structures are formed
(even at $\varepsilon_0 = 0$). The specific values of critical concentration $N_\mathrm{c}$ (critical
temperature $T_\mathrm{c}$) are governed by the elastic constants of a material, by the variation of the crystal
volume per one defect, and by the temperature (i.e., average defect concentration).

\begin{figure}[!b]
\centerline{
\includegraphics[width=0.49\textwidth]{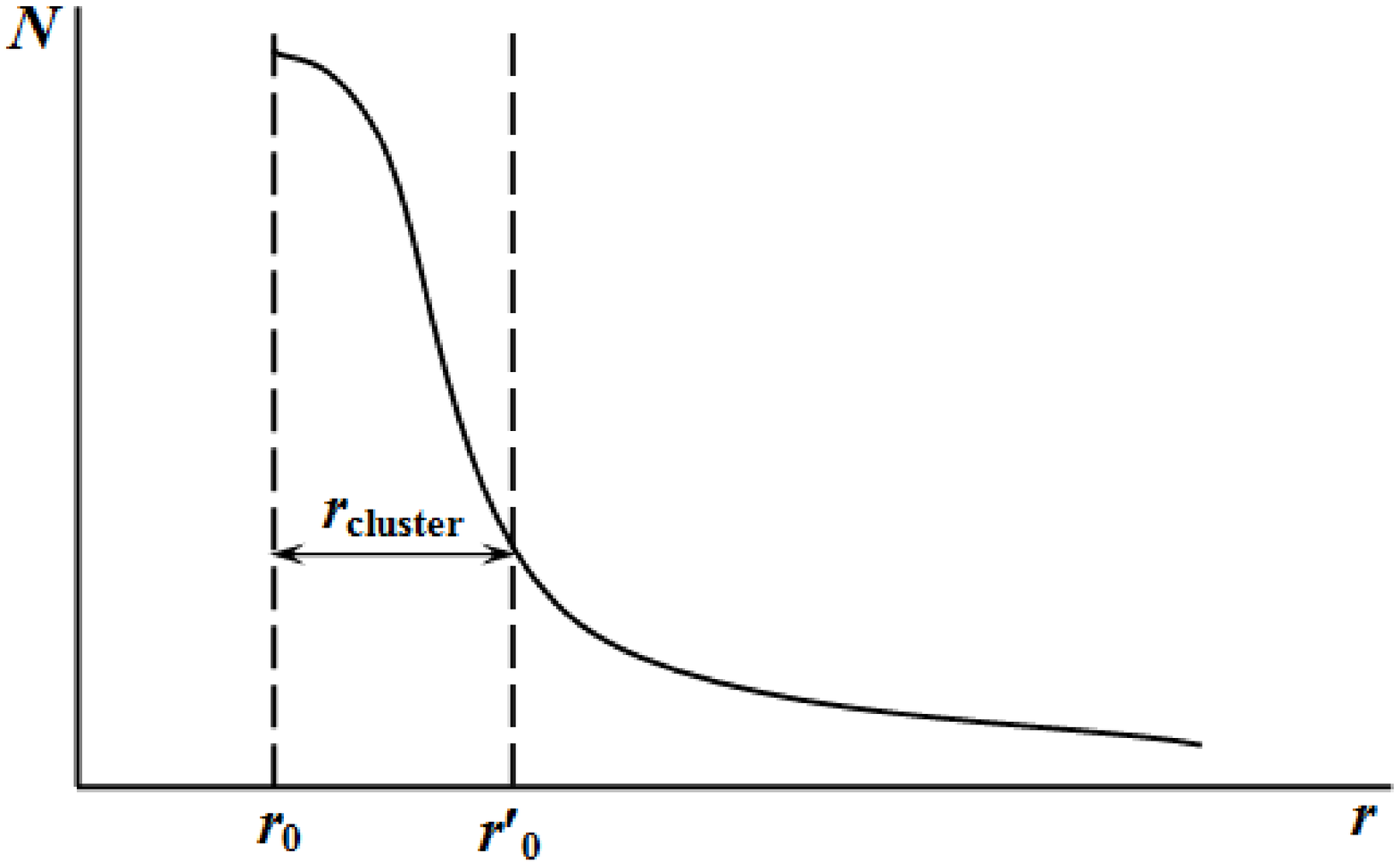}
\includegraphics[width=0.49\textwidth]{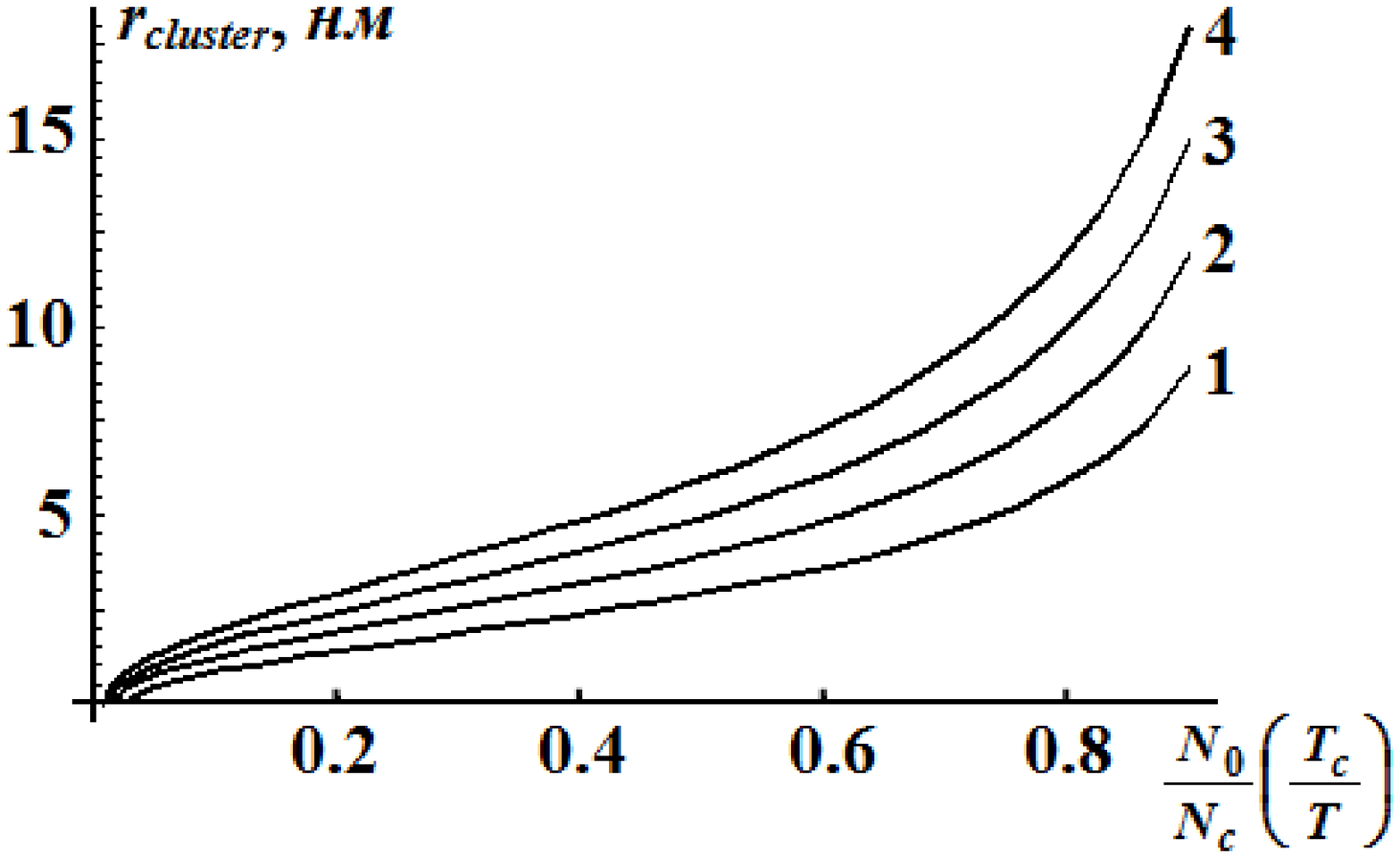}
}
\parbox[t]{0.5\textwidth}{
\caption{Spatial vacancy concentration distribution in the vicinity of the pore.}\label{fig3}
}
\hfill
\parbox[t]{0.5\textwidth}{
\caption{The dependence of the radius of a cluster of vacancies on their concentration:
1~--- $r_\mathrm{da}=3$~{nm};   2~--- $r_\mathrm{da} = 4$~{nm};   3~--- $r_\mathrm{da} = 5$~{nm};
4~--- $r_\mathrm{da} = 6$~{nm}.\label{fig4}}
}

\end{figure}

Substituting the formulae (\ref{eq23})--(\ref{eq25}) into (\ref{eq17}), we can find the vacancy concentration.
Figure~\ref{fig3} qualitatively shows the spatial vacancy concentration distribution (when a symmetric cluster is formed)
in the vicinity of a pore having the radius $r_0$.

The cluster radius depends on the defect concentration, the elastic constants, and the temperature, and can be determined as follows:
\begin{equation}
\label{eq26}
r_\mathrm{cluster}  = \frac{1}{{\sqrt a }} = r_\mathrm{da}
\sqrt {\left({{\frac{{N_0 }}{{N_\mathrm{c} }} - \frac{{a_0^2 }}{{r_\mathrm{da}^2 }}}}\right)
\left({{1 - \frac{{N_0 }}{{N_\mathrm{c} }}}}\right)^{-1}}
\end{equation}
or
\begin{equation}
\label{eq27}
r_\mathrm{cluster}  = \frac{1}{{\sqrt a }} = r_\mathrm{da}
\sqrt {\left({{\frac{{T_\mathrm{c} }}{T} - \frac{{a_0^2 }}{{r_\mathrm{da}^2 }}}}\right)\left({{1
- \frac{{T_\mathrm{c} }}{T}}}\right)^{-1}}.
\end{equation}

In figure~\ref{fig4}, the dependence of the radius of a cluster of vacancies on their relative concentration
${{N_0 }}/{{N_\mathrm{c} }}$ [temperature   $\left( {{{T_\mathrm{c} }}/{T}} \right)$] in the range
${{a_0^2 }}\big/{{r_\mathrm{da}^2 }} < {{N_0 }}/{{N_\mathrm{c} }} < 1$ is presented.  At an increase of
the vacancy concentration (i.e., a decrease of temperature), the cluster radius increases monotonously and lies in a nano-range.

\section{Diffusion coefficient}

Since vacancy clusters or their periodic structures are formed when the vacancy concentration
exceeds a certain critical value (at the temperature that is lower than a certain critical value),
we may assume that the porosity of the structure will increase due to the pore expansion and the
formation of new ones. The size  ${r'_0}$ of the pore at whose surface a vacancy cluster is formed
can be determined in the following way (see figure~\ref{fig3}):
\begin{equation}
\label{eq28}
{r'_0  = r_0  + r_\mathrm{cluster} \,.}
\end{equation}

Let us find the dependence of the diffusion coefficient $D$ of the impurities of a porous
structure on the pore radii within the the kinetic theory which is based on the assumption
that the size of impurities is much less than the distance between the impurities and between
the granules. This approximation is fairly accurate for a structure with a considerable degree
of porosity and with a small concentration of impurities.

Diffusion coefficient of gases (impurities):
\begin{equation}
\label{eq29}
D_0  = \frac{{\bar \upsilon _\mathrm{d} ^2 }}{{3z}}\,,
\end{equation}
where  $\bar \upsilon _\mathrm{d}$   is the arithmetic average velocity of the impurities;
$z$ is the number of collisions of an impurity with other impurities and granules of a porous structure per unit of  time.

\begin{wrapfigure}{i}{0.5\textwidth}
%\begin{figure}
\centerline{
\includegraphics[width=0.49\textwidth]{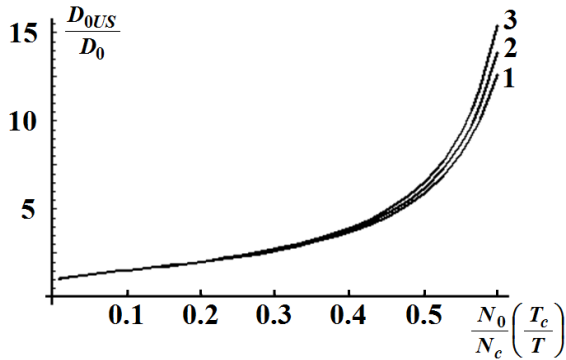}
}
\vspace{3mm}
%\parbox[t]{0.5\textwidth}{
\caption{The dependence of the relative change of a coefficient of diffusion ${{D_{0\mathrm{US}} }}/{{D_0 }}$
on the concentration of vacancies (i.e., temperature) at different radii of an impurity:
1~--- $r_\mathrm{d} = 0.45$~{nm};   2~--- $r_\mathrm{d} = 0.3$~{nm};   3~--- $r_\mathrm{d} = 0.15$~{nm}.
\label{fig5}}
%\end{figure}
\end{wrapfigure}
We define the number of collisions of an impurity with other impurities and granules of
a porous structure as the sum of the collisions with impurities and the collisions with granules taken separately.

For this purpose, we assume that the impurities and granules are globules having the
radii $r_\mathrm{d}$ and $(R_0 - r_0)$, respectively (figure~\ref{fig1}). Taking into
account that the other impurities are also moving while the granules are motionless,
the full number of the impacts can be presented as follows:
\begin{equation}
\label{eq30}
z = 4\sqrt 2 \pi r_\mathrm{d}^2 n_0 \bar \upsilon _d
+ \pi \left( {r_\mathrm{d}  + R_0  - r_0 } \right)^2 n_\mathrm{g} \bar \upsilon _\mathrm{d}\,,
\end{equation}
where  $n_\mathrm{g}$,   $n_0$ are, respectively, the granule and impurity concentrations at the granule-free bulk.
Here, it is considered that the average velocity of the relative motion of an impurity is $\sqrt 2$ times larger
than the velocity of an impurity taking into consideration the immobile granules.

The concentration of impurities at the granule-free bulk can be calculated through the concentration of
impurities in the full bulk of the structure  $n$ as
$n_0  = n{{R_0^3 }}\big/\big[{{R_0^3  - \left( {R_0^{}  - r_0^{} } \right)^3 }}\big]$.
Then, the diffusion coefficient can be written as follows:
\begin{equation}
\label{eq31}
D_0  = \frac{{\bar \upsilon _\mathrm{d} }}{{3\left[ {4\sqrt 2 \pi r_\mathrm{d}^2 n\frac{{R_0^3 }}{{R_0^3
- \left( {R_0  - r_0 } \right)^3 }}
+ \frac{3}{{4R_0^3 }}\left( {r_\mathrm{d}  + R_0  - r_0 } \right)^2 } \right]}}\,.
\end{equation}

Taking into account (\ref{eq28}), we obtain the dependence of the diffusion coefficient on the vacancy cluster size:
\begin{equation}
\label{eq32}
D_{0\mathrm{US}} = \frac{{\bar \upsilon _\mathrm{d} }}
{{3\left[ {4\sqrt 2 \pi r_\mathrm{d}^2 n\frac{{R_0^3 }}{{R_0^3  - \left( {R_0  - r_0
- r_\mathrm{cluster} } \right)^3 }} + \frac{3}{{4R_0^3 }}\left( {r_\mathrm{d}  + R_0
- r_0  - r_\mathrm{cluster} } \right)^2 } \right]}}\,.
\end{equation}

In figure~\ref{fig5}, the dependence of the relative change of a coefficient of diffusion  ${{D_{0\mathrm{US}} }}/{{D_0 }}$
on an average concentration of the vacancies (i.e., temperature) at different radii of an impurity is presented.

Such a dependence shows a monotonously increasing character. In particular, at an increase of the relative
concentration of the vacancies at ${{N_0 }}/{{N_\mathrm{c} }} = 0.55$, the coefficient of diffusion increases 10 times.

The obtained results are in good agreement with the experimental results. In particular, in~\cite{Abr94},
it is established that upon ultrasonic processing, the concentration of vacancies in solids non-linearly
depends on the amplitude of an ultrasonic wave and temperature. At a temperature below some critical value
and at a particular ultrasonic power, a significant increase of the defects of samples (more than an order
of magnitude) is observed. Thus, the acoustic and thermal effects have a pronounced synergetic character~\cite{Abr94}.
In~\cite{Kul78}, it is experimentally established that when the amplitude of an ultrasonic wave exceeds some critical
value in nickel, the pores are formed. Furthermore, at a temperature of $T < 600$\textcelsius, a significant increase of
a diffusion coefficient of carbon in nickel (by $2\div11$ times) is observed. At a temperature of $T > 600$\textcelsius,
the diffusion coefficient does not change upon ultrasonic processing.

\section{Conclusions}

\begin{enumerate}
  \item A nonlinear diffusion deformation model is presented for the formation of vacancy
nanoclusters and their periodic structures in ultrasonically treated porous material,
and the formation criteria are determined according to the deformation amplitude value,
the average vacancy concentration and temperature.
  \item Within the above mentioned model, it is shown that the diffusion coefficient of
porous structures significantly increases  at a temperature lower than some critical value,
which turns out to be in good agreement with the empirical results.
\end{enumerate}

\ukrainianpart

\title{Нелінійна модель дифузії домішок у поруватих матеріалах після ультразвукової обробки}
\author{Р.М.~Пелещак, О.В.~Кузик, О.О.~Даньків}
\address{Дрогобицький державний педагогічний університет ім. Івана Франка, \\ вул. І. Франка, 24,  82100 Дрогобич, Україна}

\makeukrtitle

\begin{abstract}
\tolerance=3000%
Створено нелінійну теорію дифузії домішок у поруватих матеріалах після ультразвукової обробки.
Показано, що при  певному значенні амплітуди деформації, середньої концентрації вакансій та
температури в результаті впливу ультразвуку можливе формування нанокластерів вакансій та їх
періодичних утворень в поруватих матеріалах. Показано, що при температурі, меншій за деяке
критичне значення, у поруватих структурах спостерігається значне зростання коефіцієнта дифузії.
\keywords коефіцієнт дифузії, ультразвук, вакансія, пора

\end{abstract}


\begin{thebibliography}{99}
\bibitem{Bul08} Bulavin~L.A., Aktan~O.Yu., Zabashta~Yu.F., Phys. Solid State, 2008,  \textbf{50}, 2270;
\doi{10.1134/S106378340812007X}.

\bibitem{Abr94} Abramov~O., Ultrasound in Liquid and Solid Metals, CRC Press, Boca Raton, 1994.

\bibitem{Eme96} Emel'yanov~V.I., Panin~I.M.,  Laser Phys., 1996, \textbf{6}, 971.

\bibitem{Sik74} Sikka~V.K., Moteff~J., J. Nucl. Mater., 1974, \textbf{54}, 325; \doi{10.1016/0022-3115(74)90144-5}.

\bibitem{Cha76} Chadderton~L.T., Johnson~E., Wohlenberg~T., Phys. Scripta, 1976, \textbf{13}, 127; \doi{10.1088/0031-8949/13/2/012}.

\bibitem{Zav02} Zaveryukhin~B.N., Zaveryukhina~N.N., Tursunkulov~O.M., Tech. Phys. Lett., 2002, \textbf{28}, 752; \doi{10.1134/1.1511774}.

\bibitem{Oli02} Olikh~O.Ya., Ostrovskii~I.V., Phys. Solid State, 2002, \textbf{44}, 1249; \doi{10.1134/1.1494617}.

\bibitem{Ost95} Ostapenko~S., Bell~R., J. Appl. Phys., 1995, \textbf{77}, 5458; \doi{10.1063/1.359243}.

\bibitem{Ost02} Ostrovski~I.V., Nadtochi~A.B., Podolyan~A.A., Semiconductors, 2002, \textbf{36}, 367; \doi{10.1134/1.1469179}.

\bibitem{Kul78} Kulemin~A.V., Ultrasound and Diffusion in Metals, Metallurgiya, Moscow, 1978  (in Russian).

\bibitem{Kun70} Kunin~I.A., Nonlocal Theory of Elasticity, Polish Academy of Sciences, Warsaw, 1970.

\end{thebibliography}
\end{document}